\documentclass[aps,twocolumn,superscriptaddress,showpacs,prl]{revtex4}
\usepackage{psfrag}
\usepackage{amssymb}
\usepackage{graphicx}
\def\be{\begin{equation}}
\def\ee{\end{equation}}
\def\bea{\begin{eqnarray}}
\def\eea{\end{eqnarray}}

\begin{document}
\title{Purity estimation with separable measurements}
\author{E.~Bagan}
\affiliation{Grup de F{\'\i}sica Te{\`o}rica \& IFAE, Facultat de Ci{\`e}ncies,
Edifici Cn, Universitat Aut{\`o}noma de Barcelona, 08193 Bellaterra
(Barcelona) Spain}
\author{M.~A.~Ballester}
\affiliation{Department of Mathematics, University of Utrecht, Box
80010, 3508 TA Utrecht, The Netherlands}
\author{R.~Mu{\~n}oz-Tapia}
\affiliation{Grup de F{\'\i}sica Te{\`o}rica \& IFAE, Facultat de Ci{\`e}ncies,
Edifici Cn, Universitat Aut{\`o}noma de Barcelona, 08193 Bellaterra
(Barcelona) Spain}
\author{O.~Romero-Isart}
\affiliation{Grup de F{\'\i}sica Te{\`o}rica \& IFAE, Facultat de Ci{\`e}ncies,
Edifici Cn, Universitat Aut{\`o}noma de Barcelona, 08193 Bellaterra
(Barcelona) Spain}
%\date{\today}
%
\begin{abstract}
Given a large number $N$ of copies of a qubit state of which we wish to estimate its purity,
we prove that
separable-measurement protocols can
be as efficient as the optimal joint-measurement  one if classical
communication is used. This shows that the optimal
estimation of the entanglement of a two-qubit state  can also be
achieved asymptotically with fully separable measurements. Thus, quantum memories
provide no advantage in this situation.
The relationship between our global Bayesian approach and the quantum
Cram\'er-Rao bound is discussed.
\end{abstract}
 \maketitle
The ultimate goal of quantum state estimation is to determine the
value of the parameters that fully characterize a given unknown
quantum state. However, in practical applications,
a partial characterization %of an unknown state
is often all one needs. Thus, e.g., knowing the purity of a qubit
state or the degree of entanglement of a bipartite state may be
sufficient to determine whether it can perform some particular
task~\cite{white} ---See Ref.~\cite{gisin} for recent experimental
progress on estimating the degree of polarization (the purity) of
light beams. This Letter concerns this type of situation.

To be more specific, assume we are given $N$ identical copies of
an unknown qubit mixed state $\rho(\vec r)$, so that the state of
the total system is $\rho^N(\vec r)\equiv[\rho(\vec r)]^{\otimes
N}$. The set of all such density matrices $\{\rho(\vec r)\}$ can
be mapped into the Bloch sphere ${\cal B}=\{\vec r :\ r\equiv|\vec
r|\le1\}$ through the relation $\rho(\vec r)=(\openone+\vec
r\cdot\vec\sigma)/2$, where
$\vec\sigma=(\sigma_x,\sigma_y,\sigma_z)$ is a vector made out of
the three standard Pauli matrices. Our aim is to estimate the purity, $r$, as
accurately as possible by performing suitable measurements on the
$N$ copies, i.e., on $\rho^N(\vec r)$. This problem can also be
viewed as the parameter estimation of a depolarizing
channel~\cite{depolarizing} when it is fed with $N$ identical states.

Estimation protocols are broadly divided into two classes
depending on the type of measurements they use: joint and
separable. The former treats the system of $N$ qubits as a whole,
allowing for the most general measurements, and leads to the most
accurate estimates or, equivalently, to the largest fidelity
(properly defined below). The latter, treats each copy separately
but classical communication can be used in the measurement
process. This class is particularly important because it is
feasible with current technology and it offers an economy of
resources. In this Letter we show that for a sufficiently large
$N$, separable measurement protocols for purity estimation can
attain the  optimal joint-measurement fidelity bound. The power of
separable measurement protocols in achieving optimal performance
has also been demonstrated in other contexts~\cite{others}.

It has been shown~\cite{vidal} that given $N$ copies of a
bipartite qubit pure state, $|\Psi\rangle_{AB}$, the optimal
protocol for measuring its entanglement consists in estimating the
purity of $\rho(\vec r)\equiv{\rm
tr}_B(|\Psi\rangle_{AB}\langle\Psi|)$, where ${\rm tr}_B$ is the
partial trace over the Hilbert space of party $B$
(see~\cite{horodecki} for related  work on bipartite mixed
states). We thus show that for {\em large $N$} this entanglement
can be optimally estimated by performing just {\em separable}
measurements on {\em one} party (party $A$ in this discussion) of
{\em each} of the $N$ copies of~$|\Psi\rangle_{AB}$.

In this Letter, special
attention is paid to  the asymptotic regime, when $N$ is large.
There are several reasons for this. First, in this limit, formulas
greatly simplify and usually reveal important features of the
estimation protocol. Second, the asymptotic theory of quantum
statistical inference, which has become in recent years a very
active field in mathematical statistics~\cite{masahito-book},
deals with  problems such as the one at hand. Our results give
support to some quantum statistical methods for which only
heuristic proofs exist; e.g., the applicability  of the integrated
quantum Cram\'er-Rao bound in the Bayesian approach~\cite{us-prep}.

In the first part of this Letter we state some important results concerning the optimal joint
estimation protocols and give the corresponding fidelity bounds. In
addition to the general case of states in $\cal B$, which was
partially addressed in~\cite{vidal}, we also consider the situation
when the unknown state is constrained  to lie on the equatorial
plane $\cal E$ of the Bloch sphere $\cal B$. In the second part,
we discuss separable measurement protocols, we prove that they
saturate the joint-measurement bound asymptotically and we state our
conclusions.

Mathematically, the problem of estimating the purity of $\rho(\vec
r)$ can be formulated within the Bayesian framework as follows
(see~\cite{keyl} for a large deviations approach, which is only
meaningful in the asymptotic regime). Let ${\cal R}_{\cal
O}=\{R_\chi\}$ be the set of estimates of $r$, each of them  based
on a particular outcome $\chi$ of some generalized measurement,
$\cal O$, over $\rho^N(\vec r)$. Such measurement is characterized
by a Positive Operator Valued Measure (POVM), namely, by a set of
positive operators ${\cal O}=\{O_\chi\}$ that satisfy $\sum_\chi
O_\chi=\openone$. A separable measurement is a particularly
interesting instance of a POVM for which each $O_\chi$ is a tensor
product of $N$ individual operators (usually projectors) each one
of them acting on $\rho(\vec r)$.

Next, a figure of merit, $f(r,R_\chi)$, is introduced as a
quantitative way of expressing the quality of the purity
estimation. Throughout this Letter we use
\begin{equation}
f(r,R_\chi)=rR_\chi+\sqrt{1-r^2}\sqrt{1-R_\chi^2},
\label{fidelity}
\end{equation}
which we call fidelity for short. Its values are in the range
$[0,1]$, where unity corresponds to perfect determination. This fidelity has
a natural interpretation:  in Uhlmann's geometric representation
of the set of density matrices as the hemisphere $(1/2){\mathbb
S}^3\subset{\mathbb R}^4$, the function $D(r,R_\chi)=(1/2)\arccos
f(r,R_\chi)$ is the geodesic (Bures) distance~\cite{som} between
two sets  (two parallel 2-dimensional spheres) characterized by
the purities~$r$ and~$R_\chi$ respectively.

The optimal protocol is obtained by maximizing
$F({\cal O},{\cal R}_{\cal O})=\sum_\chi\int d\rho f(r,R_\chi) {\rm
tr}[\rho^N(\vec r) O_\chi]$,
where $d\rho$ is the prior probability distribution of $\rho(\vec
r)$, and we identify the trace as the probability of obtaining the
outcome $\chi$ given that the state we measure upon is
$\rho^N(\vec r)$. Thus, $F$ is the average fidelity. The
maximization is over the estimator (guessed purity) ${\cal
R}_{\cal O}$ and the POVM ${\cal O}$,
\begin{equation}
F^{\rm max}\equiv \max_{\cal O}\left\{\max_{{\cal R}_{\cal O}}F({\cal O},{\cal
R}_{\cal O})\right\} .
\label{optimal fidelity}
\end{equation}

In this formulation, we need to provide a prior probability
distribution (prior for short) $d\rho$, which encodes our initial
knowledge about $\rho(\vec r)$. Here we assume to be completely
ignorant of both $\vec n\equiv \vec r/r$ and $r$. Our lack of knowledge about the
former is properly  represented with the choice $d\rho \propto
d\Omega$ (solid angle element), which states that {\em \`a priori}
$\vec n$ is isotropically distributed on ${\cal B}$. Therefore, we
write
\begin{equation}
d\rho={d\Omega\over4\pi} w(r)dr;\quad \int_0^1dr\, w(r)=1.
\label{measure}
\end{equation}
While there is  wide agreement on this respect, the $r$-dependence
of the prior is controversial and so far  we will not stick to any
particular choice.
Nevertheless, it is worth keeping in mind that  the hard sphere prior $w(r)=3 r^2$ shows
up in the context of entanglement estimation~\cite{zycz},
whereas the Bures
prior $w(r)=(4/\pi) r^2 (1-r^2)^{-1/2}$ is most
natural in connection with distinguishability of density matrices~\cite{fuchs,fid,prior}.

Rather than computing~(\ref{optimal fidelity}), we here present the main results
(details can be found in~\cite{us longer}).

({\em i}\hspace{.15em})~The optimal POVM is defined by the set of
operators $\{\openone_{j\alpha}=\sum_m|jm;\alpha\rangle\langle
jm;\alpha|\}$. Each $\openone_{j\alpha}$ projects over the
invariant subspace corresponding to an irreducible representation
$\bf j$ of $SU(2)$ ---the group of unitary transformations $U$
that acts naturally over $\cal B$ as $\rho(\vec r)\to U\rho(\vec
r)U^\dagger$. Here the index $\alpha$ ($\alpha=1,2,\dots, n_j$)
labels the different $n_j$ occurrences of $\bf j$. All these $n_j$
equivalent representations $\bf j$ give an identical contribution
to $F$. This result should not come as a surprise. The optimal
purity estimate, and thus the fidelity, should only depend on
invariant quantities (i.e., $j$ and $\alpha$), as the purity
itself  is rotationally invariant, and so is our choice for the
prior.

({\em ii}\hspace{.15em})~The optimal purity estimator can be written as
$R_j=A_j(A_j^2+B_j^2)^{-1/2}$, where
\begin{equation}
\kern-0.25em (A_j,B_j)=\!\!\int dr\,w(r)\,(r,\sqrt{1-r^2})\!\!\sum_{m=-j}^j\!\! p_r^{{N\over2}-m}
q_r^{{N\over2}+m},
\label{Vjchi}
\end{equation}
and $p_r=(1-r)/2$, $q_r=1-p_r$. We can easily  identify the sum
in~(\ref{Vjchi}) as the probability ${\rm tr}[\rho^N(\vec
r)\openone_{j\alpha}]$.

({\em iii}\hspace{.15em})~The maximum fidelity is given by
\begin{equation}
F^{\rm
max}=\pmatrix{N\cr{N\over2}-j}{2j+1\over{N\over2}+j+1}\sum_j\sqrt{A_j^2+B_j^2} \ , \label{Fmax}
\end{equation}
where the coefficient in front of the sum
is~$n_j$~\cite{cirac,us-prep}.

For large $N$, this can be computed~to~be~\cite{us-prep}
\begin{equation}
F^{\rm max}= 1-{1\over2N}+ o(N^{-1}) . \label{Fasymp}
\end{equation}
One can also check that at leading order $R_j=2j/N+\dots$, as one would intuitively expect.
These asymptotic results hold for any prior $w(r)$.

It is also interesting to analyze the case where $\vec r$
is known to lie on the equatorial plane $\cal E$. With
this information, the prior probability distribution
becomes $d\rho=(d\phi/2\pi)w(r)dr$, where $\phi$ is the
polar angle of the spherical coordinates.  The group of
unitary transformations on~$\cal E$ is now a $U(1)$
subgroup of $SU(2)$ and, hence, the optimal POVM is given
by the corresponding one-dimensional projectors over the
$U(1)$-invariant subspaces, $\{\openone_{j\alpha
m}\equiv|jm;\alpha\rangle\langle jm;\alpha|\}$.  With
this, one can work out the maximum fidelity. It turns out
that asymptotically $F^{\rm max}$ is also given by~(\ref{Fasymp}) and the
optimal guess is $R_{jm}=2j/N+\dots$ (independently of
$m$). The same conclusions also hold in the
one-dimensional case of states known to lie on a diameter
of $\cal B$. Therefore, we see that the information
about~$\vec n$ becomes irrelevant in the asymptotic limit.

A word regarding quantum statistical inference is in order here.
It is often argued that the quantum Cram\'er-Rao
bound~\cite{holevo} can be integrated to provide an attainable
asymptotic lower bound for some averaged figures of merit, such as
the fidelity~(\ref{fidelity}).  Ours is a so-called one parameter
problem for which the quantum Cram\'er-Rao bound takes the simple
form ${\rm Var}\, R\ge H^{-1}(\vec r)/N$, where  ${\rm Var}\,
R\equiv\langle (R_\chi-\langle R_\chi\rangle)^2\rangle$ is the
variance of the estimator~$R_\chi$, the average is over the
outcomes $\chi$ of a measurement,  $H(\vec r)$ is the quantum
information matrix~\cite{holevo}, and $R_\chi$ is assumed to be
unbiased: $ \langle R_\chi\rangle=r $. In our case $H(\vec
r)=(1-r^2)^{-1}$, and the bound is attainable. This provides in
turn an attainable asymptotic upper bound for the
fidelity~(\ref{fidelity}), since $\langle
f(r,R_\chi)\rangle\approx
1-\raisebox{.12em}{\mbox{\tiny$1\over2$}}H(\vec r)\,{\rm Var}\,
R+\dots$. Assuming one can integrate  this relations over the
whole of~$\cal B$ (including the region $r\approx1$, where $H(\vec
r)$ is singular), with a weight function given by the
prior~(\ref{measure}), one easily obtains Eq.~(\ref{Fasymp}).
Unfortunately, there are only heuristic arguments supporting this
assumption, but so far no rigorous proof exists in the
literature.

%%%%%%%%%

We now abandon the joint protocols to dwell on separable-measurement
strategies for the rest of the Letter.  Here we focus
on the asymptotic regime, but some brief comments concerning small
$N$ can be found in the conclusions.

In previous work~\cite{alberto}, some of the authors showed that
the maximum fidelity one can achieve in estimating both $r$ and
$\vec n$ (full estimation of a qubit mixed state) assuming the
Bures prior and using tomography
behaves as
$
F^{\rm max}_{\rm full}=1-{\xi\, N^{-3/4}}+o(N^{-3/4})   ,
$
where $\xi$ is a positive constant. The same behavior one should
expect for our fidelity $F^{\rm max}$, since the effect of the
purity estimation is dominant in~$F^{\rm max}_{\rm full}$. This strange power
law, somehow unexpected on statistical grounds, is caused by the
behavior of $w(r)$  in a small region \mbox{$r\approx 1$}. Indeed,
it is not difficult to convince oneself that if $w(r) \propto
(1-r^2)^{-\lambda}\approx 2(1-r)^{-\lambda}$ for $r\approx 1$, one
should have $1-F^{\rm max}\propto N^{\lambda/2-1}+\dots$, for
$0<\lambda<1$ (for $\lambda=0$, hard sphere prior, one should
have  logarithmic corrections). This differs drastically
from~(\ref{Fasymp}) which, as stated above, holds for
{\rm any} such values of $\lambda$. Would classical communication
be enough to restore the right power law $N^{-1}$ for $1-F^{\rm
max}$ and, moreover, saturate the bound of the optimal joint-measurement protocol?

On quantum statistical grounds, one should expect a positive
answer to this question since  the quantum Cram\'er-Rao bound is
attained by a separable protocol consisting in performing the (von
Neumann) measurements ${\cal M}=\{(\openone\pm\vec
n\cdot\sigma)/2\}$ on each copy. Note, however, that $\cal M$
depends on $\vec n$, which is, of course, unknown {\em
\`a priori}. This protocol can only make sense if we are ready to
spend a fraction of the $N$ copies of $\rho(\vec r)$ to obtain an
estimate of~$\vec n$, use this classical information to design
$\cal M$ and, finally, perform this adapted measurement on the
remaining copies. This protocol was successfully applied to pure
states by Gill and Massar in~\cite{gill-massar}. We extend it to
purity estimation below.

Let us consider a family of priors of the form
\begin{equation}
w(r)={4\over\sqrt\pi}{\Gamma(5/2-\lambda)\over\Gamma(1-\lambda)}{r^2
(1-r^2)^{-\lambda}}  , \label{gen prior}
\end{equation}
which includes both the Bures ($\lambda=1/2$) and the hard sphere ($\lambda=0$) metrics. Despite
of this particular $r$ dependence, the final results apply to any
prior whose behavior near $r=1$ is given by~(\ref{gen prior}).

We now proceed {\em \`a la} Gill-Massar~\cite{gill-massar} and
consider the following one-step adaptive protocol: we take a
fraction $N^\alpha\equiv N_0$ ($0<\alpha<1$) of the $N$ copies
of $\rho(\vec r)$ and we use them to estimate $\vec n$. Tomography
along the three orthogonal axis $x$, $y$ and $z$, together with a
very elementary estimation based on the relative frequencies of
the outcomes~\cite{us-local},  enables us to estimate $\vec n$
with an accuracy given by
\begin{equation}
{\langle\Theta^2_r\rangle\over
2}\approx1-\langle\cos\Theta_r\rangle={3\over N_0}\left({1\over
r^2}-{1\over5}\right)+o(N_0^{-1}), \label{Theta}
\end{equation}
where $\Theta_r$ is the angle between $\vec n$ and its estimate.
Here and below $\langle\cdots\rangle$ is not only the average over
the outcomes of this tomography measurements, but also contains an
integration over the prior angular distribution $d\Omega/(4\pi)$
for fixed $r$.

In a second step, we measure the projection of $\vec\sigma$ along
the estimated $\vec n$ obtained in the previous step. We perform
this von Neumann measurement on each of the remaining $N-N_0\equiv
N_1$ copies of the state $\rho(\vec r)$. We estimate the purity to
be $R=2N_+/N_1-1$,  where $N_\pm/N_1$ is the relative frequency of
$\pm1$ outcomes, and we drop the $N_+$ dependence of $R$ to
simplify the notation.

Obviously, as a random variable and for large $N_1$, $R$~is normally
distributed as $R\sim{\rm N}(r c_r,\sqrt{1-r^2
c^2_r}/\sqrt{N_1})$, where $c_r=\cos\Theta_r$. Hence, for large
$N_0$ and $N_1$ it makes sense to expand $f(r,R)$,
Eq.~(\ref{fidelity}), around $R= r c_r$,  and thereafter, because
of~(\ref{Theta}), expand the resulting expression around~$c_r=1$.
We obtain
\begin{equation}
F(r)= 1-{1\over 2
N_1}+{r^2\over1-r^2}\left({\langle\Theta^2_r\rangle\over4N_1}-{\langle\Theta^4_r\rangle\over8}\right)+\dots
, \label{<f>}
\end{equation}
where $F(r)$ is the average fidelity for fixed $r$, i.e., $\int
dr\,w(r) F(r)=F$. In view of~(\ref{Theta}),
$\langle\Theta_r^4\rangle\sim N_0^{-2}=N^{-2\alpha}$. Hence, the
two terms in parenthesis in~(\ref{<f>}) can only be dropped if
$\alpha>1/2$. Provided $w(r)$ vanishes as in~(\ref{gen prior})
with $\lambda<0$, we can integrate $r$ in~(\ref{<f>})  over the
unit interval to obtain
\begin{equation}
F=1-{1\over2N(1-N^{\alpha-1})}+o(N^{-1}) , \label{F in I}
\end{equation}
and we conclude that this protocol attains asymptotically the
joint-measurement bound~(\ref{Fasymp}).

However, most of the physically interesting priors~\cite{prior,zycz}, $w(r)$, not
only do not vanish as $r\to1$, but often diverge like~(\ref{gen
prior})  with $0<\lambda<1$. In this case (\ref{<f>}) cannot be
integrated, as the last term does not lead to a convergent
integral. This signals that the series expansion around $c_r=1$
leading to~(\ref{<f>}) is not legitimated in the whole of~$\cal
B$.

To fix the problem, we split $\cal B$ in two regions. A sphere
of radius $1-\epsilon$, $\epsilon>0$, which we call ${\cal B}^{\rm
I}$, and a spherical sheet  of thickness $\epsilon$:  ${\cal
B}^{\rm II}=\{\vec r: 1-\epsilon<r\le 1\}$. The fidelity can thus
be written as the sum of the corresponding two contributions:
$F=F^{\rm I}+F^{\rm II}$. While $F^{\rm I}$ can be obtained by
simply integrating~(\ref{<f>}) over ${\cal B}^{\rm I}$, where this
expansion is valid, some care must be taken in the region~${\cal B}^{\rm II}$. There, we proceed
as follows.

We compute the fidelity as if all the states in ${\cal B}^{\rm II}$ had the lowest possible purity ($r=1-\epsilon$)
when the first-step tomography was performed. This leads to a lower bound for $F^{\rm II}$,
because the lower the purity of a state  the less accurately  $\vec n$ can be determined [see Eq.~(\ref{Theta})],
and hence, the worse its purity can be estimated in the
second step.
The trick, which amounts to replacing $c_r$ by $c_{1-\epsilon}$, enables us to integrate~$r$ prior
to performing the average~$\langle\cdots\rangle$.  A~straightforward calculation leads to
\begin{equation}
F\gtrsim 1-{1\over2N_1}-2^{\lambda-2}k_\lambda
\langle\Theta_{1-\epsilon}^2\rangle^{2-\lambda} +\dots,
\label{fidelity ok}
\end{equation}
$0<\lambda<1$, where $
k_\lambda=2^{2-\lambda}\Gamma({5\over2}-\lambda)\Gamma({3\over2}-\lambda)\Gamma(\lambda-2)/[\pi\Gamma(1-\lambda)]
$.
Now, we can safely take the limit
$\epsilon\to0$. We see that by choosing
$
{\rm
\max}\left\{{1/2},{1/(2-\lambda)}\right\}<\alpha< 1
$
 we ensure that the joint-measurement
bound~(\ref{Fasymp}) is attained. It is worth
emphasizing that the last term in~(\ref{fidelity ok}), which is
completely missing in~(\ref{F in I}), is actually the dominant
contribution if $\alpha<1/(2-\lambda)$. For $\lambda=0$ we have
$F^{\rm hard}\gtrsim1-{(1/2)N_1^{-1}}-{(3/8) N_1^{-1}  \langle\Theta^2_1\rangle
\log\langle\Theta^2_1\rangle}+\dots  $
and we again conclude that the protocol presented here attains the
joint-measurement bound.

%%%%%

At this point one may wonder if the conclusions above depend upon
our particular choice of figure of merit. To get a grasp on this,
it is worth using again the standard pointwise approach to quantum
statistics. There, one is interested in the mean square error
${\rm MSE}\,R=\langle(R-r)^2\rangle$ for fixed $r$, where now the
average $\langle\cdots\rangle$ is over the outcomes of {\em all}
measurements for a fixed $\vec r$. One can write ${\rm
MSE}\,R={\rm Var}\, R + (\langle R\rangle-r)^2 $, where the second
term is the \emph{bias}. Using the same one-step adaptive protocol
described above, we get that the mean square error after step two
is $ {\rm MSE}\,R=[N_1H(r)]^{-1}+r^2\langle\Theta_r^4\rangle/4+
\dots $. As above, the last term can be dropped if $\alpha>1/2$,
and
${\rm MSE}\,R=[N \,H(r)]^{-1}+o[N^{-1}]$,
saturating the quantum Cram\'er-Rao bound. This protocol is,
therefore, also asymptotically optimal in the present context.

%%%%

In summary, though the absolute bounds for the average fidelity
involve joint measurements, these bounds can be obtained
asymptotically with separable measurements. This requires
classical communication among the sequential von Neumann
measurements performed on each of the $N$ individual copies of the
state. This result, which has been speculated on quantum
statistical grounds, is here proved for the first time by a direct
calculation. Since the purity is an optimal measure of the
entanglement  of a  pure bipartite qubit  state, we also obtain
the additional result that this entanglement can be optimally
estimated with separable measurements on just one of the parties.

For finite (but otherwise arbitrary) $N$, finding the optimal
separable measurement protocol is an open problem. Interestingly
enough, a `greedy' protocol designed to be optimal at each
measurement step~\cite{others} leads to an unacceptably poor
estimation.  Notice that in the one-step adaptive  protocol
described above, part of the copies were spent (`wasted'  from a
`greedy'  point of view) in estimating $\vec n$. We have seen that
this strategy pays in the long run. However, the `greedy' strategy
optimizes measurements in the short run, which translates into
measuring $\vec\sigma$ along the same arbitrarily fixed axis.
This yields a low value for the fidelity, which does not
even converge to unity in the strict limit $N\to\infty$.
This counterintuitive behavior
also appears in
other contexts as, e.g., economics, biology or social sciences
(see e.g.~\cite{parrondo}).

We thank Antonio Ac{\'\i}n, Richard Gill and Juanma Parrondo for useful
discussions. This work is supported by the  Spanish Ministry of
Science and Technology project BFM2002-02588, CIRIT project
SGR-00185, Netherlands Organization for Scientific Research NWO,
the European Community projects QUPRODIS  contract no.
IST-2001-38877 and RESQ contract no. IST-2001-37559.

%%%%%%%%%%%%%%%%%%%%%

\newcommand{\PRL}[3]{Phys.~Rev. Lett.~\textbf{#1}, #2~(#3)}
\newcommand{\PRA}[3]{Phys.~Rev. A~\textbf{#1}, #2~(#3)}
\newcommand{\JPA}[3]{J.~Phys. A~\textbf{#1}, #2~(#3)}
\newcommand{\PLA}[3]{Phys.~Lett. A~\textbf{#1}, #2~(#3)}
\newcommand{\JOB}[3]{J.~Opt. B~\textbf{#1}, #2~(#3)}
\newcommand{\JMP}[3]{J.~Math.~Phys.~\textbf{#1}, #2~(#3)}
\newcommand{\JMO}[3]{J.~Mod.~Opt.~\textbf{#1}, #2~(#3)}

\end{document}